\documentclass[
journal=arXiv, 
manuscript=article]{achemso}

\usepackage[version=3]{mhchem} 
\usepackage{hyperref}

\author{Davide Nardone}
\author{Angelo  Ciaramella}
\affiliation[Parthenope University]
{Dept. of Science and Technology, University of Naples Parthenope}
\author{Mariangela Cerreta}
\affiliation[IBP]
{The Institute of Protein Biochemistry (IBP) of the Italian National Research Council (CNR)}
\author{Salvatore Pulcrano}
\author{Gian Carlo Bellenchi}
\affiliation[IGB]
{The Institute of Genetic and Biophysic (IGB) of the Italian National Research Council (CNR)}
\author{Giuseppe Manco}
\author{Ferdinando Febbraio}
\affiliation[IBP]
{The Institute of Protein Biochemistry (IBP) of the Italian National Research Council (CNR)}

\title[\texttt{achemso} demonstration]
{SELYMATRA: Web Application for the analysis of mass spectra}

\begin{document}

\begin{abstract}
\noindent
\textbf{Motivation:} Surface Enhanced Laser Desorption/Ionization-Time Of Flight Mass Spectrometry (SELDI-TOF MS) is a variant of the MALDI. It is uses in many cases especially for the analysis of protein profiling and for preliminary screening tasks of complex sample aimed for the searching of biomarker. Unfortunately, these analysis are time consuming and strictly limited about the protein identification. Seldi analysis of mass spectra (SELYMATRA) is a Web Application (WA) developed with the aim of reduce these lacks automating the identification processes and introducing the possibility to predict the proteins present in complex mixtures from cells and tissues analysed by Mass Spectrometry.
\\
\noindent
\textbf{Results:} SELYMATRA has the following characteristics. The architectural pattern used to develop the WA is the \textit{Model-View-Controller} (MVC), extremely used in the development of software system. The WA expects an user to upload data in a \textit{Microsoft Excel spreadsheet file} format, usually generated by means of the proprietary Mass Spectrometry softwares. Several parameters can be set such as experiment conditions, range of isoelectric point, range of pH, relative errors and so on. The WA compare the mass value among two mass spectra (sample \textit{vs} control) to extract differences, and according to the parameters set, it queries a local database for the prediction of the most likely proteins related to the masses differently expressed. The WA was validated in a cellular model overexpressing a tagged NURR1 receptor.
\\
\noindent
\textbf{Availability and Implementation:} \href{}{SELYMATRA is available at\\ http://140.164.61.23:8080/SELYMATRA}
\\
\noindent
\textbf{Contact:} \href{}{selymatra@ibp.cnr.it}
\end{abstract}

\section{Introduction}

Comparison of the protein profiling in cellular systems could give important information about the changes in the protein expression after cell alterations (such as in oxidative stress or in cancer) so as to better understand the cell answer to this events as well as for the discovery of signals disease to be used as diagnostic markers or targets for drugs. SELDI-TOF is a variant of the MALDI mass spectrometry especially used for the high throughput analysis of protein profile in cell samples. This technique exploits active matrices to select/sort proteins on the basis of their chemical-physical characteristics (\citealp{Chung L}). Despite the high potential of this technique, the lack in protein identification and the difficulty to analyze complex spectra reduce the fields of application. Here, we present a WA named SELYMATRA, which has the aim of significantly reducing the analysis time of a large amount of data and at the same time to obtain information on proteomic profiles so far not obtainable. The architectural pattern used to develop the WA is the \textit{Model-View-Controller} (MVC), extremely used in the development of software system.

\section{Methodology and Results}

\begin{figure}[t!]
  \includegraphics[scale=0.45]{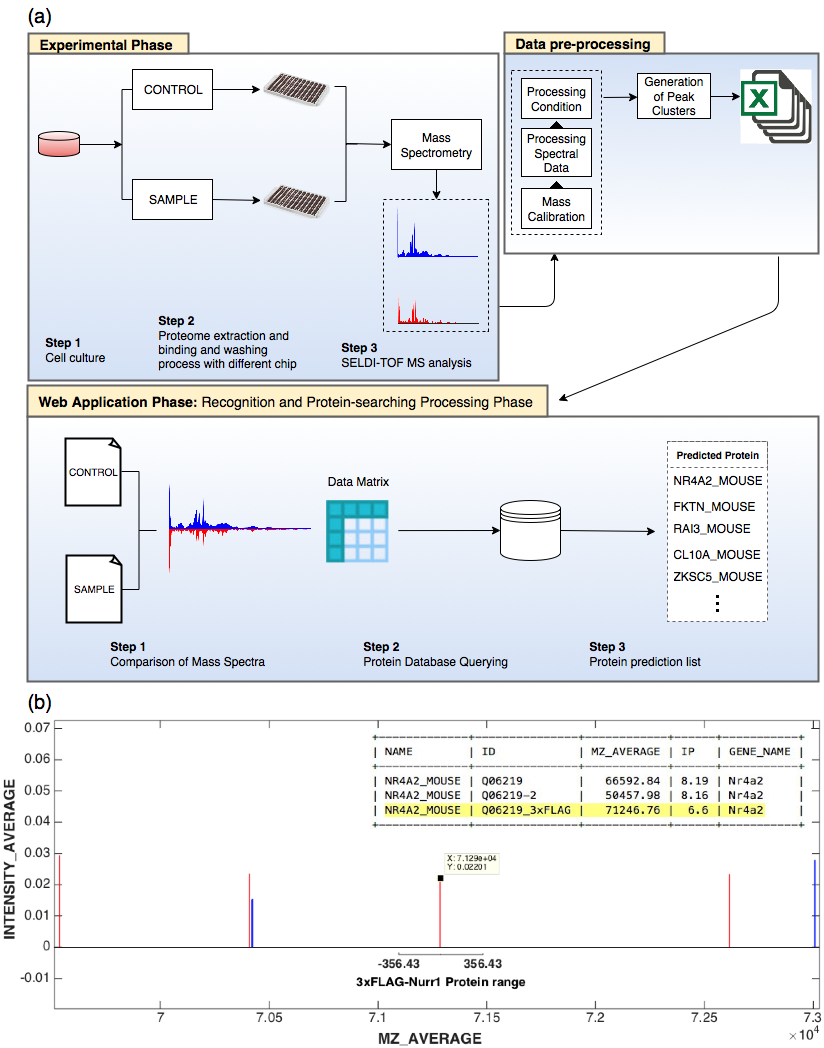}
\caption{\textbf {Workflow overview of SELYMATRA. A)} {MS data experimentally obtained from replicate measurements of whole protein extracts, from cytoplasm and nuclei of mouse A1 dopaminergic cell lines, whose were pre-processed in clusters using ProteinChip Data Manager application. The output data-sheets were used by SELYMATRA for the comparison of clustered peaks and the identification of the up- and down-regulated proteins. By accessing to a local mouse protein database, (together with parameters from the experimental conditions), a prediction of proteins ID for each mass peak were also done. \textbf{B)} MS clustering of protein samples extracted from nuclei of mouse A1 dopaminergic cells (in blue) and over-expressing 3\textsuperscript{x}FLAG-Nurr1 (in red). In the insert, the SELYMATRA results for the identification of 3\textsuperscript{x}FLAG-Nurr1, with a mw of about 72 KDa.}}
\label{fig:ex1}
\end{figure}

The WA aims to carry out a particular pattern recognition analysis, looking for specific similarities among two mass spectra under different washing and binding conditions, (process that's not just feasible using the SELDI software packet) and in doing so, it yields to a deeper data analysis which reduce significantly the lifetime of some particular experiments (Fig.1A). Once a file is uploaded, the WA applies a method of analysis and processing data, consisting mainly of:
\begin{enumerate}
\item \textit{Mass spectra comparison and recognition process};
\item \textit{Protein ID searching process.}
\end{enumerate}
Respectively, through these two processes the WA elaborates the data to perform a \textit{key feature extraction} among two mass spectra and give differences as well as up- and down-regulation in protein expressions. According to these results and to the characteristics of the experiment, it queries a local database for the \textit{prediction of the most likely proteins} related to the previous results (Fig.\ref{fig:ex1}A).
Regarding the \textit{Mass spectra comparison and recognition process} different results may be obtained depending on the choice and values of some parameters. Basically, each procedure is developed in a different way, but the underlying idea is quite similar. That's to say, a \textit{math inequality} is used for each modeling-procedure, by which a result-set of values is generated. It's defined as follow:
\begin{equation}
(m/z\_avg-m/z\_std)\ast\delta<\textbf{x}<(m/z\_avg+m/z\_std)\ast\delta\label{eq:01}
\end{equation}
where \textbf{m/z\_avg} and \textbf{m/z\_std} are respectively the mean value and the standard deviation of the sample (A), {\boldmath$\delta$} is a correction factor used for increasing or diminishing the range searching, and \textbf{x} is the current \textit{m/z value} to find in the sample (B) for the inequality~(\ref{eq:01}). 

\clearpage
According to the user needs, four type of m/z can be identified for any experiment:
\begin{itemize}
\item \textbf{found:} set of \textit{m/z\_avg} satisfying the Equation~(\ref{eq:01});
\item \textbf{missing:} set of \textit{m/z\_avg} not satisfying the Equation~(\ref{eq:01}) or rather that represent the complement of the \textit{found} (A) masses against to the all (B) masses, expressed formally as: A$\setminus$B = $B-A$ = $\lbrace$x$\in$B $\wedge$ x$\notin$A$\rbrace$;
\item \textbf{upgrade or downgrade:} set of \textit{m/z\_avg} satisfying the Equation~(\ref{eq:01}) and the \textit{median intensity of mass variation constrict}, expressed as:
\begin{equation}
\frac{x-y}{y}*100\leq\lambda\label{eq:02}
\end{equation}
\end{itemize}
where \textbf{x} and \textbf{y} are respectively the \textit{m/z\_avg} of the searching sample (B) and the current one (A) (which both satisfy the Equation~(\ref{eq:01})) and, {\boldmath$\lambda$} is an index of \textit{variation of the intensity median} for distinguish the affinity among the relative intensities of two mass peaks.\\
Since a set of identified mass obtained by the \textit{mass spectra comparison and recognition process} is not an exhaustive result, we have built-up a process to infer more information on such masses, in particular predicting the most likely protein bound to them. In order to make this possible, for each mass of the \textit{result-set} has been considered:

\begin{itemize}
\item A dynamic amplification factor (depending on the m/z value) used for the assessment of a \textit{molecular mass (Mw) error-range}; 
\item a type of chip and binding/washing condition for the assessment of an \textit{isoelectric point (pI) range}.
\end{itemize}
Both these ranges allow to identify a confidence interval in which the \textit{protein searching process} takes place. As it has been stated initially, this process is possible through the building, maintaining and tuning of a database of proteins, whose information have been retrieved through java API provided by Uniprot database (\citealp{Magrane M}).
Moreover by introducing this last process, the WA is able to recognize rough patterns of proteins related to the phenomena represented into the target sample.
An example was the identification, in nuclei of A1 cell line, of the transcription factor 3\textsuperscript{x}FLAG-Nurr1 (\citealp{Volpicelli F}), over-expressed in the expression cassette TET-ON (Fig.\ref{fig:ex1}B). As expected, the SELYMATRA WA has associated the single MS peak at about 72 KDa to the 3\textsuperscript{x}FLAG-Nurr1 ID artificially added to the local database (Fig.1B in yellow).
Although interesting outcomes may be obtained by this process, the number of proteins retrieved may be very cumbersome (effect depending not to a wrong calculation but rather to the MW and pI's constraints range definition we stated in the \textit{protein searching process}). As consequence, a new method of \textit{proteins selection} can be developed in the next work in order to discriminate and cut-out irrelevant proteins.\\
The data elaborated by SELYMATRA are strongly related to the SELDI-TOF MS or MALDI's technology and their own software (e.g., ProteinChip Data Manager) which have many functionality such as the \textit{identification of peaks}, its \textit{clustering} and also the intensity comparing among two mass spectra. Regardless of their usefulness, these software are not able to classify cluster of peaks having different intensity values as well as to identify potential outliers, which instead is possible through SELYMATRA.\\

\section{Conclusion}
The development of this WA has led to the acquisition of new results not attainable earlier. Particularly, through ad-hoc modeling procedures it's possible to obtain \textit{key features} representing potential outliers or features discerning the similarity among two mass spectra. Based on this result it has also been possible to predict sets of proteins, simply interfacing the WA to a proteome database. As result of some achievements, such WA has significantly made easier the research work conducted on the analysis of protein expression in the differentiation of dopaminergic cell line A1 driven by 3\textsuperscript{x}FLAG-Nurr1 over-expression, so as to speeding up the lifetime data analysis and the processing of the experimental results, by doing so of the WA a worthy tool of interest for the experts in biological field.

\section*{Acknowledgement}
\textbf{Funding:} The research leading to these results has received funding from FIRB: Medical Research in Italy (MERIT). Grant n.2: ref. 0017153.

\end{document}